\documentclass{article}

\usepackage{microtype}
\usepackage{graphicx}
\usepackage{subcaption}
\usepackage{booktabs} 
\usepackage{multirow}

\usepackage{hyperref}

\usepackage[accepted]{icml2026}

\usepackage{amsmath}
\usepackage{amssymb}
\usepackage{mathtools}
\usepackage{amsthm}
\usepackage{makecell}
\usepackage{booktabs}   
\usepackage{amssymb}    
\usepackage{pifont}     
\usepackage{tabularx}
\newcommand{\cmark}{\ding{51}} 
\newcommand{\xmark}{\ding{55}} 
\usepackage[table]{xcolor}
\usepackage[capitalize,noabbrev]{cleveref}

\theoremstyle{plain}

\theoremstyle{definition}

\theoremstyle{remark}

\usepackage[textsize=tiny]{todonotes}

\icmltitlerunning{Non-Canonical AMP Discovery with AMPGAN v3}

\begin{document}

\twocolumn[
  \icmltitle{Agentic Discovery of Non-Canonical Antimicrobial Peptides with AMPGAN v3}



  \begin{icmlauthorlist}
    \icmlauthor{Jay Hwasung Jung}{UVM}
    \icmlauthor{Xiaohan Zhang}{UVM}
    \icmlauthor{Shenghan Song}{PUMC}
    \icmlauthor{Mahmoud Sayedahmed}{PUCP}
    \icmlauthor{Chijian Xiang}{PUMC,PUHLA}
    \icmlauthor{Yunong Xu}{PUMC}
    \icmlauthor{Ahmed AbdelKhalek}{PUCP}
    \icmlauthor{Severin T. Schneebeli}{PUIMP}
    \icmlauthor{Matthew J. Wargo}{UVMMC}
    \icmlauthor{Jianing Li}{PUMC}
    \icmlauthor{Safwan Wshah}{UVM}
  \end{icmlauthorlist}

  \icmlaffiliation{UVM}{Department of Computer Science, University of Vermont, Burlington, VT, USA}
  \icmlaffiliation{UVMMC}{Department of Microbiology and Molecular Genetics, Larner College of Medicine, University of Vermont, Burlington, VT, USA}
  \icmlaffiliation{PUMC}{Borch Department of Medicinal Chemistry and Molecular Pharmacology, Purdue University, West Lafayette, IN, USA}
  \icmlaffiliation{PUCP}{Department of Comparative Pathobiology, Purdue University, West Lafayette, IN, USA}
  \icmlaffiliation{PUHLA}{Department of Horticulture and Landscape Architecture, Purdue University, West Lafayette, IN, USA}
  \icmlaffiliation{PUIMP}{Department of Industrial and Molecular Pharmaceutics, Purdue University, West Lafayette, IN, USA}
  \icmlcorrespondingauthor{Jianing Li}{jianing-li@purdue.edu}
\icmlcorrespondingauthor{Safwan Wshah}{Safwan.Wshah@uvm.edu}
  \icmlkeywords{Antimicrobial Peptides (AMPs), GAN, De novo Peptide Design, Agentic AI, Multi-agent systems, Generative Models for Biology}

  \vskip 0.3in
]



\printAffiliationsAndNotice{}  

\begin{abstract}
Antimicrobial resistance causes to over a million deaths annually. Antimicrobial peptides (AMPs) are a promising solution, but generative AMP models are not yet ready to design peptides with non-natural amino acids and/or chemical modifications, which are essential for real-world peptide drugs. We present \textbf{AMPGAN v3}, a multi-objective conditional GAN that expands the generative vocabulary to D-amino acids and N/C-terminus modifications such as amidation. By separating adversarial and activity-aware supervision across two specialized discriminators, AMPGAN~v3 substantially improves training stability and outperforms prior generative AMP models on external classifiers. We validated five candidates spanning three structural classes \textit{in vitro}; two showed activity against Gram-positive strains, with the best candidate reaching MIC 8 $\mu$g/mL against \textit{B. subtilis}. To support downstream curation, we further present \textbf{PepCraft}, a multi-agent framework for end-to-end AMP discovery in which a Planning Agent orchestrates specialized executors for generation, filtering, and verification. Its prioritization recommendations align with our in vitro outcomes. Together, these contributions let us examine, on a small but real scale, how generative and agentic AI compose in therapeutic peptide discovery. Code: \href{https://github.com/marszzibros/AMPGANv3}{https://github.com/marszzibros/AMPGANv3}.
\end{abstract}

\section{Introduction}
\label{sec:intro}
\begin{table*}[ht!]
\caption{Comparison of generative methods for antimicrobial peptide (AMP) discovery. AMPGAN v3 is the only method supporting D-amino acids and terminal modifications, significantly expanding the accessible chemical space beyond canonical L-amino acids.}
\label{tab:amp_comparison}
\centering
\small
\setlength{\tabcolsep}{5pt}
\renewcommand{\arraystretch}{1.2}

\begin{tabular}{l c c c c c c }
\toprule
\textbf{Method} & \textbf{Architecture} & 
\makecell{\textbf{Conditional}\\\textbf{Generation}} & 
\makecell{\textbf{Multi-}\\\textbf{Objective}} & 
\makecell{\textbf{Species}\\\textbf{Targeting}} & 
\makecell{\textbf{Chemical}\\\textbf{Modifications}} & 
\makecell{\textbf{in vitro}\\\textbf{Validation}}  \\

\midrule

PepGAN \cite{pepgan} & GAN & \xmark & \xmark & \xmark & \xmark & \cmark \\
HydrAMP \cite{hydramp} & cVAE & \cmark & \xmark & \xmark & \xmark & \cmark  \\
HMAMP \cite{HMAMP} & GAN (multi-D) & \cmark & \cmark & \xmark & \xmark & \cmark \\
MOFormer \cite{moformer} & Transformer & \cmark & \cmark & \xmark & \xmark & \xmark \\
AMP-Designer \cite{AMP_GPT} & LLM (GPT) & \cmark & \cmark & \cmark & \xmark & \cmark \\
AMPGAN v2 \cite{ampganv2} & cGAN & \cmark & \xmark & \xmark & \xmark & \cmark$^*$ \\

\midrule
\textbf{AMPGAN v3 (Ours)} & \textbf{cGAN (multi-D)} & 
\cmark & \cmark & \cmark & \cmark & \cmark  \\

\bottomrule
\end{tabular}

\vspace{2pt}
\begin{flushleft}
\scriptsize \textbf{Note:} \cmark = supported, \xmark = not supported, $^*$AMPGAN v1~\cite{ampgan_bio} includes \textit{in vitro} validation

\end{flushleft}
\vspace{-10pt}

\end{table*}
Despite decades of pharmaceutical investment, only five novel classes of antibiotics have been commercialized since 2000, a period often referred to as the \textit{``discovery void''}~\cite{discovery_void}. This slump is attributed to the high cost and low commercial returns of antibiotic development~\cite{antibiotics_investment}. Most importantly, the emergence of antimicrobial resistance (AMR) continues to hinder the efficacy of current therapeutics~\cite{WHO}, contributing to 1.27 million global deaths in 2019~\cite{AMR_kills_millions}. Addressing this escalating health crisis requires both new antimicrobial therapeutics with anti-resistance mechanisms and scalable approaches to discover them. 

Antimicrobial peptides (AMPs) are a leading candidate: relatively short sequences (typically $<$ 64 amino acids) that act through multiple mechanisms, often disrupting bacterial membranes via non-specific physical interactions,a mechanism associated with slower resistance emergence~\cite{AMP_membranes}. AMPs are particularly suitable to data-driven \textit{de novo} design because substantial experimentally-validated database exist~\cite{dbaasp,AMP_fact}, and standard activity assays are sufficiently inexpensive to enable wet-lab validation of generated candidates. 

These properties have motivated generative AI approaches for AMP discovery. Early work leveraged variational autoencoders~\cite{hydramp} and adversarial architectures~\cite{pepgan,ampgan_bio,ampganv2} to produce AMP-like sequences. Subsequent studies introduced multi-objective optimization approaches~\cite{HMAMP,moformer}, while most recent methods adopt large language model (LLM)-based foundation models with pretraining and prompt tuning~\cite{AMP_GPT}. Despite this progress, two limitations persist across these approaches: (1) \textit{In vitro} tested AMPs are predominately $\alpha$-helical. Short helical peptides suffer from conformational instability and rapid proteolytic degradation, limiting their therapeutic viability~\cite{alpha_helical_instable}. (2) Existing methods omit chemical modifications (e.g., D-amino acids, terminal modifications) from their training. In real-world therapeutic applications, these modifications are indispensable; peptides without them are highly susceptible to degradation in minutes before showing any therapeutic effect~\cite{peptide_need_modifications}. 

Among prior approaches, AMPGAN~\cite{ampgan_bio,ampganv2} is the work most directly related to ours: a conditional GAN that supports target-microbe conditioning. However, AMPGAN suffers from severe training instability, only $\sim$10\% of random initializations produce usable models, with failed runs collapsing to repeated single-token outputs~\cite{GAN_collapse}. Its conditioning is also restricted to L-amino acid peptides of 32 residues or fewer, showcasing the chemical-scope limitations described above.

Improving the generative model addresses only part of the challenge. Even if a model generates highly stable sequences, translating those \textit{in-silico} designs into experimentally testable candidates requires filtering against physicochemical constraints, and cross-referencing against known databases. These repetitive steps scale poorly with increased generative output volume when performed manually. Recent work in LLM-based scientific agents has begun to explore how such workflows can be orchestrated with minimal human intervention. While agentic systems have been applied to chemistry~\cite{autolab} and protein engineering~\cite{proteincrow,bioreason_pro} they remain unexplored for AMP discovery. Unlike scripted pipelines, agentic systems support adaptive control flow: the Planning Agent can re-invoke executors with revised plans when intermediate results deviate from the user objective. By combining controllable generation with specialized executor agents, we explore how generative and agentic AI can be composed in an end-to-end AMP discovery pipeline.

In this work, \textbf{we present AMPGAN v3}, a multi-objective conditional generative model that jointly optimizes the sequence realism and antimicrobial activity for AMP discovery. AMPGAN v3 expands the chemical scope of prior approaches to include D-amino acids, and terminal modifications, and produces structurally diverse outputs spanning $\alpha$-helical, $\beta$-hairpin, and random coil conformations, enabling the generation of peptides with chemical stability required for therapeutic viability. By separating adversarial training from activity prediction across two specialized discriminators, we resolve the training instability reported in AMPGAN v2, significantly improving successful-run rates. We validate AMPGAN v3 through \textit{in vitro} experiments, testing five generated candidates spanning $\alpha$-helical, $\beta$-hairpin, and random-coil conformations and incorporating D-amino acids and amidation, which were not previously supported by AMP generative models. Two of five candidates exhibited clear antimicrobial activity against Gram-positive species. Finally, we develop an exploratory multi-agent framework for end-to-end AMP discovery, which we refer to as \textbf{PepCraft}, demonstrating it with AMPGAN v3 as the generative component. A Planning Agent orchestrates specialized executors for sequence generation, physicochemical filtering and database verification, exploring how generative and agentic AI can be composed in scientific discovery pipelines.

\section{Background and Related Work}
\label{sec:background}

\subsection{Antimicrobial Peptides (AMPs)}
AMPs are short sequences of fewer than 100 amino acids that exhibit wide ranges of activity against bacteria, fungi, and viruses by primarily targeting microbial membranes~\cite{AMP_membranes,AMP_fact}. They are generally cationic (+1 to +9) and amphipathic, facilitating electrostatic attraction to negatively charged microbial membranes and subsequent insertion into the lipid bilayer. Accordingly, AMPs commonly adopt secondary structures such as $\alpha$-helices and $\beta$-sheets, which support their amphipathic architecture~\cite{AMP_chemical_fact}. AMPs offer several advantages over conventional antibiotics: they often act through multiple targets, reducing the likelihood of resistance development~\cite{AMP_membranes}, and they exhibit rapid bactericidal activity with membrane disruption occurring within seconds to minutes~\cite{AMP_fast}. These properties make AMPs particularly attractive candidates against multidrug-resistant (MDR) pathogens. 



\subsection{Generative Models for AMP Discovery}
\label{sec:generative_models}

Generative models have illustrated great promise in discovering active antimicrobial peptides. Early approaches focused on balancing sequence fidelity with biological activity. PepGAN~\cite{pepgan} employs a generative adversarial network guided by an activity predictor to filter out inactive sequences. Similarly, HydrAMP~\cite{hydramp} utilizes a conditional Variational Autoencoder (VAE) to disentangle peptide representations from antimicrobial conditions. This approach allows for both unconstrained and analogue generation. Subsequent research expanded into multi-objective optimization and large foundation models. HMAMP~\cite{HMAMP} trains a generator against multiple discriminators to optimize distinct attributes like minimum inhibitory concentration and hemolysis. However, it relies on reward-based gradient estimation, which suffers from non-differentiable limitations. MOFormer~\cite{moformer} addresses multi-objective design using a conditional Transformer with Pareto-based feedback. Taking a different route, AMP-Designer~\cite{AMP_GPT} adapts a large language model architecture. It pretrains on vast peptide databases to achieve high \textit{in vitro} and \textit{in vivo} efficacy. For controllable design, AMPGAN v2 \cite{ampganv2} adopted bidirectional conditional GANs. It conditions peptide generation on specific target microbes and mechanisms. The generated peptides were highly diverse and yielded a high predicted-active rate. A comprehensive comparison of these methods is provided in~\cref{tab:amp_comparison}.



\subsection{Agentic Systems in Computational Biology}
Large Language Models (LLMs) are now driving autonomous agentic systems. These systems orchestrate tools, reason over data, and refine outputs. As a result, they are reshaping computational biology and wet-lab workflows. Recent works such as AutoLabs~\cite{autolab} break down complex experiments into tasks for specialized, self-correcting agents. It translates natural language into hardware-ready protocols for liquid-handling robots. Similarly, ProteinCrow~\cite{proteincrow} applies this approach directly to protein engineering. It equips an LLM with curated tools for structure prediction and inverse design, allowing autonomous execution that previously requires extensive manual oversight. BioReason-pro~\cite{bioreason_pro} demonstrates that LLMs equipped with bioinformatics tools can produce natural-language interpretations of biological sequences. 


\section{Materials and Methods}
\label{sec:methodology}
\subsection{\textit{In Vitro} Protocol}
To validate the biological activity of the generated antimicrobial peptide candidates, selected peptides were synthesized and tested using standard MIC and MBC assays. Peptides were dissolved in sterile water to prepare 10 mg/mL stock solutions. Bacterial strains used for evaluation are summarized in~\cref{app:invitro}. Strains were cultured overnight at 37 \textdegree C under aerobic conditions using tryptic soy medium, except for Enterococcus faecium, which was cultured using brain heart infusion medium. Bacterial cultures were diluted to approximately $5\times 10^5$CFU/mL before peptide treatment.

MIC values were determined using a two-fold serial dilution assay in 96-well plates. Peptide dilutions were mixed with bacterial suspensions and incubated at 37 \textdegree C for 16 h. The MIC was defined as the lowest peptide concentration that resulted in no visible bacterial growth. To determine MBC values, samples from wells without visible growth were plated onto antibiotic-free agar and incubated under the same conditions. The MBC was defined as the lowest peptide concentration that prevented bacterial regrowth. MIC and MBC results are summarized in~\cref{tab:mic_mbc}.

\subsection{AMPGAN-v3 Model Architecture}

\begin{figure*}[ht!]
    \centering
    \includegraphics[width=1.0\textwidth]{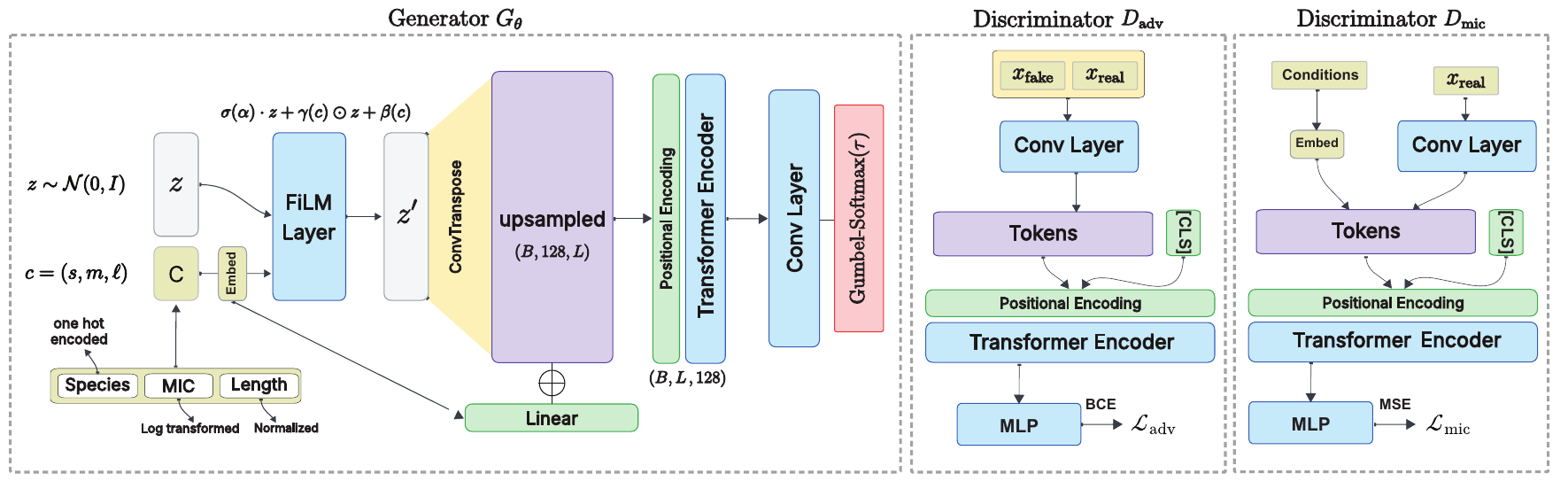}
    
    \caption{\textbf{AMPGAN-v3 architecture.} The generator $G_\theta$ maps a latent vector $z \sim \mathcal{N}(0, I)$ and condition $c = (\text{species} - s, \text{MIC} - m, \text{length}-\ell)$ to a sequence of token logits via FiLM-based conditioning, transposed convolutions, and a Transformer encoder. Two discriminators are trained jointly: $D_\text{adv}$ classifies real vs. generated sequences unconditionally, whereas $D_\text{mic}$ regresses MIC values from real sequences conditioned on species.  }
    \label{fig:ampganv3}
\vspace{-10pt}
\end{figure*}
We formulate conditional AMP generation as a multi-objective adversarial learning task. Given a training set $\mathcal{D} = \{x_i, c_i\}_{i=1}^{N}$, where $x_i \in \{0,1\}^{L\times V}$ is a one-hot encoded peptide sequence of maximum length $L$ over a vocabulary of $V$ tokens, and $c_i =(s_i, m_i,\ell_i)$ is a condition tuple comprising target species $s_i$, minimum inhibitory concentration (MIC) $m_i$, and peptide length $\ell_i$. We train a generator $G$ against two specific discriminators: $D_{\text{adv}}$ for adversarial training and $D_{\text{mic}}$ for antimicrobial activity regression. The overall architecture of AMPGANv3 is shown in~\cref{fig:ampganv3}.

\textbf{Generator}: The generator $G_\theta: \mathcal{Z} \times \mathcal{C} \rightarrow \mathbb{R}^{L \times V}$ maps a latent vector $z \sim \mathcal{N}(0, I_{d_z})$ and condition $c=(s, m, \ell)$ to a sequence of token logits. Each condition is embedded into a shared space $\mathbb{R}^{d_e}$: species $s$ via a learned embedding table, and the MIC value $m$ and normalized length $\ell$ via learned linear projections. The combined condition vector is their concatenation $\mathbf{c} = [\mathbf{e}_s ; \mathbf{e}_m ; \mathbf{e}_\ell] \in \mathbb{R}^{3d_e}$. We modulate the latent code via Feature-wise Linear Modulation (FiLM)~\cite{film} with gating: $z' = \sigma(\alpha)\cdot z+\gamma(c)\odot z+\beta(c)$ where $\gamma, \beta$ are learned affine transformations of c and $\alpha$ is a learnable gating scalar. The modulated vector is progressively upsampled through transposed convolutions to length $L$, then refined by a Transformer encoder. A final convolutional stack projects to logit space $\mathbb{R}^{L\times V}$.

\textbf{Discriminators}: We separate adversarial and activity-aware supervision into two discriminators rather than coupling them into a single conditional Discriminator. First, the two signals answer different questions: $D_\text{adv}$ asks whether a sequence is plausibly peptide-like, a distributional judgment over the full training corpus, whereas $D_\text{mic}$ asks whether a sequence is consistent with a specified potency, a conditional judgment that depends on the target MIC. Both discriminators share a convolutional stem that projects the one-hot token dimension into a subspace with dimension $d$, followed by a Transformer encoder with a learnable [CLS] token prepended to the sequence. Then, the [CLS] representation is passed through an MLP head for prediction. $D_{\text{adv}}$ operates unconditionally on both real and generated sequences and outputs a binary real/fake classification. $D_{\text{mic}}$ is trained exclusively on real sequences, conditions on species by an extra embedding token to the sequence before the Transformer encoder, and regresses a MIC value. Its parameters are updated exclusively on real sequences with experimentally measured MIC labels; during generator training,$D_\text{mic}$ is frozen and scores generated sequences to provide gradients to Generator.

\textbf{Multi-Objective Training}: The generator is trained against both discriminators simultaneously. At each training step, we first update $D_{\text{adv}}$ on real and generated samples, then update $D_{\text{mic}}$ on real samples only, and finally update G on a weighted combination of four objectives:
    $\mathcal{L}_G = 
    \lambda_\ell \mathcal{L}_\text{len} + \lambda_s \mathcal{L}_\text{sim} + \lambda_a \mathcal{L}_\text{adv} + \lambda_m \mathcal{L}_\text{mic} $
where $\mathcal{L}_\text{adv}$ and $\mathcal{L}_\text{mic}$ are computed with discriminator parameters frozen. Since $G$ produces continuous logits over a discrete vocabulary, we apply the Gumbel-Softmax estimator with temperature $\tau$ to obtain differentiable discrete samples $\hat{x} = \text{Gumbel-Softmax}_\tau(G(z,c))$ for both discriminators. The adversarial term is the standard cross-entropy (CE) against $D_\text{adv}$, and the conditional term is mean-squared error (MSE) that pushes generated sequences toward the target potency $c$ as scored by $D_\text{mic}$ conditioned on species $s$:
\begin{equation}
\begin{aligned}
\mathcal{L}_\text{adv} &= -\frac{1}{B}\sum_{i} \log D_\text{adv}(\hat{x}_i), \\
\mathcal{L}_\text{mic} &= \frac{1}{B}\sum_{i} \lVert D_\text{mic}(\hat{x}_i, s_i) - c_i \rVert^2.
\end{aligned}
\end{equation}
The remaining two terms operate directly on the generator's predicted token distribution $\hat{p}_{i,t,v}$: the probability assigned to vocabulary token $v$ at position $t$ in sequence $i$. For each sample $i$, let $y_i$ be the real sequence with tokens $y_{i,t}$ at position $t$ and EOS at position $\ell_i$, and full sequence length $L_i =\ell_i+1$ accounting for a special terminus token~(See \cref{sec:dataset} for details). The length loss signals sequence termination by maximizing the predicted EOS probability at each sequence's true EOS position:
\begin{equation}
\mathcal{L}_\text{len} = -\frac{1}{B}\sum_{i} \log \hat{p}_{i,\ell_i,\text{EOS}}
\end{equation}
The similarity loss $\mathcal{L}_\text{sim}$ extends this per-position supervision across the full sequence, providing a token-level reconstruction signal that stabilizes the early training. Sequences positions are weighted at 1 and padding positions at $\alpha=0.1$, so the peptide reconstruction signal dominates while trailing tokens are mildly discouraged. 

\begin{equation}
\mathcal{L}_\text{sim} = -\frac{\sum_{i,t} w_{i,t} \log \hat{p}_{i,t,y_{i,t}}}{\sum_{i,t} w_{i,t}}, \quad w_{i,t} = \begin{cases} 1 & t \le L_i \\ \alpha & t > L_i \end{cases}
\end{equation}



\subsection{Agentic Discovery Pipeline}
\label{sec:methodology_agentic}

\begin{figure*}[ht!]
    \centering
    \includegraphics[width=1.0\textwidth]{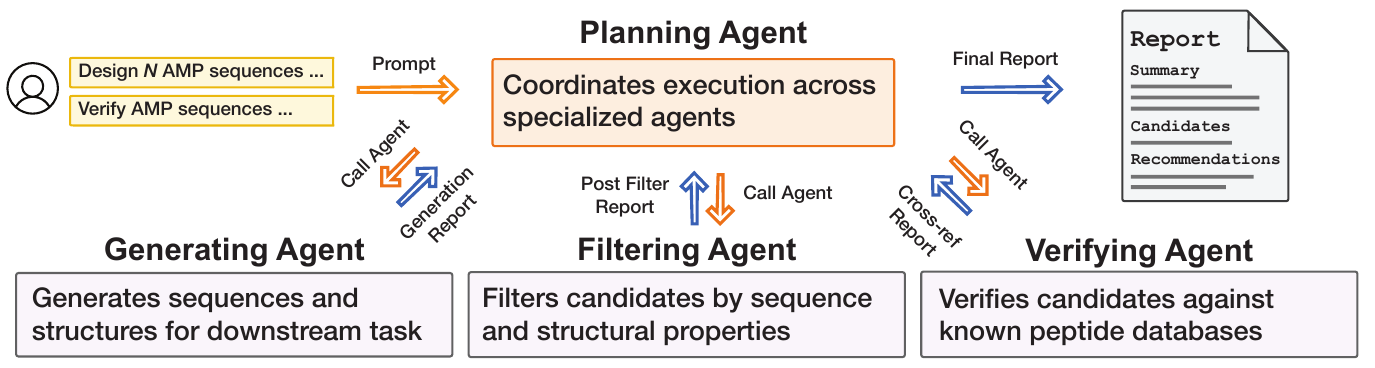}
    
    \caption{\textbf{Agentic workflow for AMP discovery.} A Planning Agent translates user objectives into natural-language instructions and send them to specialized executors with tools: Generating Agent, Filtering Agent and Verifying Agent. Each executor returns a report, and Planning Agent iterates until the objective is met.}
    \label{fig:agentic}
    \vspace{-10pt}
\end{figure*}
We propose a multi-agent framework for automated AMP design, orchestrated by a Planning Agent $\mathcal{P}$ that dynamically selects and instructs specialized executor agents based on iterative feedback as illustrated in Figure~\ref{fig:agentic}. Specifically, at each iteration $t$, the Planning Agent receives the user objective 
$\theta_{0}$ and the previous report $r_{t-1}$ where $r_{0} = \emptyset$. It then selects an executor agent $e$ and prompts it with a specific instruction $i^e_t$ at iteration $t$.

\begin{equation}
    \mathcal{P}\left(\theta_{0},\ r_{t-1}\right) = 
    \begin{cases}  
        \text{END} & \text{If Objective Fulfilled} \\
        (e, i^e_{t}) & \text{Otherwise}
    \end{cases}.
\end{equation}

The selected executor $e$ executes its designated tools along with the instruction prompt generated by the planning agent $\mathcal{P}$ and returns 
a run report $r_{t} = e(i^e_t)$. Example instructions and reports are provided in \cref{app:agentic}.

\textbf{Planning Agent}: Decomposes the user-defined objective into natural language instructions and calls the appropriate executor. Upon receiving a run report $r_{t}$, it decides whether to re-invoke an executor with revised parameters or proceed to the next stage. 

\textbf{Generating Agent.} Translates user-defined natural-language objectives into structured generation calls for AMPGAN v3 and SimpleFold~\cite{simplefold}-based structure prediction. 

\textbf{Filtering Agent.} Applies sequential physicochemical and structural screens to refine the raw candidate pool. Property-level filters (charge, hydrophobicity, length) run first, and structure-dependent filters run only on candidates that survive earlier passes. Each filtering cycle concludes with a status check that the Planning Agent uses to plan.

\textbf{Verifying Agent.} Assesses novelty and biological relevance through BLAST~\cite{blast} homology search against SwissProt~\cite{Swissprot} (broad protein context) and the AMPGAN v3 training dataset (DBAASP~\cite{dbaasp}; near-duplicate detection and known-activity recovery). Annotated hits are returned to the Planner as the basis for prioritization recommendations.

Agent system prompts, tool specifications, and a representative trajectory are provided in Appendix~\ref{app:agentic}.



\section{Dataset Curation}
\label{sec:dataset}
To enable the training of our AMPGAN-v3, we curate a dataset of experimentally validated AMP sequences collected from DBAASP v3~\cite{dbaasp} via its public REST API. We queried all peptide IDs sequentially, retaining only monomers with sequences of 64 residues or fewer. For each retained entry, we extracted the amino acid sequence, N-terminus and C-terminus modifications, and per-species antimicrobial activity, retaining only MIC and IC measurements. Concentration values reported as ranges were replaced by their midpoint, and all concentrations were converted to $\mu$g/mL. Entries with missing units or unparseable concentration values were excluded. Sequences were stored in their original case to preserve D-amino acid annotations, and the resulting dataset was deduplicated by sequence. Each sequence is encoded as a structured token sequence of the form [N-term][SOS] [Amino Acid Tokens] [EOS][C-term], where [N-term] and [C-term] denote terminus-modification tokens (or none) and [SOS]/[EOS] mark sequence boundaries. This format allows the model to generate terminus modifications and sequence content jointly within a single output stream.

The final training dataset contains 34{,}275 rows corresponding to 10{,}892 unique sequences. A single sequence may appear multiple times, each paired with a distinct combination of N-terminus modification, C-terminus modification, or target species. The six target species are \textit{E. Coli}, \textit{P. Aeruginosa}, \textit{K. Pneumoniae}, \textit{S. Aureus}, \textit{B. Subtilis}, and \textit{S. Epidermidis}. During training, each row is treated as an independent sample, allowing the model to learn the mapping from a sequence-species pair to its inhibitory activity (MIC).

In contrast to AMPGAN v2~\cite{ampganv2}, which assigned arbitrarily high MIC values to represent inactivity, our approach relies entirely on experimentally measured concentrations. Antimicrobial activity is inherently species-dependent. Because a sequence can be highly effective against one species while remaining inactive against another, presetting MIC values for untested pairs would inject misleading supervision signals into the training process.
\section{Experiments}
\label{sec:experiments}
We evaluate AMPGAN v3 along three stages: (1) \textit{in vitro} antimicrobial activity of synthesized candidates against clinically relevant bacterial strains, (2) \textit{in-silico} generation quality, comparing against AMPGAN v2~\cite{ampganv2} and published sequences from the reviewed methods~\cref{sec:generative_models}, and (3) the utility of our agentic workflow relative to single-prompt baselines. More details can be found in~\cref{app:ampgan}.

\subsection{\textit{In Vitro} Evaluation}
\label{sec:invitro}

\textbf{Candidate selection.} From AMPGAN v3 generations conditioned on \textit{E.Coli} and \textit{S.Aureus} with target MIC $<32~\mu\text{g/mL}$, we filtered candidates using four physicochemical criteria common in AMP screening: instability index $<$ 50, net charge at pH~7 between 2 and 9, GRAVY score between -1 and 1, and isoelectric point $>9$. Secondary structure was predicted with AlphaFold3~\cite{alphafold3}, and chemical modifications (D-amino acid substitutions, C-terminal amidation) were introduced using Schrödinger Maestro~\cite{maestro}. From the filtered pool, we selected five candidates spanning three structural classes: two $\alpha$-helices (HY-P60322, HY-P60323), two $\beta$
-hairpins (HY-P60324, HY-P60325), and one random coil (HY-P60326). The selected candidates and their target species are listed in \cref{tab:mic_mbc}.
 
\setlength{\fboxsep}{0.5pt}
\newcommand{\Daa}[1]{\colorbox{red!20}{#1}}
\newcommand{\species}[2]{\shortstack{\textit{#1} \\ \textit{#2}}}
\begin{table*}[ht!]
\centering
\caption{Minimum inhibitory and bactericidal concentrations (MIC/MBC, \textmu g/mL) of peptides against the bacterial panel.}
\label{tab:mic_mbc}
\setlength{\tabcolsep}{4pt}
\footnotesize
\begin{tabular}{lccccc}
\toprule

& & \species{Staphylococcus}{aureus}
& \species{Bacillus}{subtilis}
& \species{Staphylococcus}{epidermidis}
& \species{Enterococcus}{faecium} \\

Sequence code & Full sequences
& \scriptsize USA300-0114
& \scriptsize ---
& \scriptsize ATCC 35984
& \scriptsize HM-952 \\

& &
\scriptsize Wound\textsuperscript{a}
& \scriptsize ---
& \scriptsize Catheter
& \scriptsize Human, USA \\

\midrule

HY-P60323 &
KRTASKVKKH\Daa{k}RNAATIAVA\Daa{h}AR-NH$_2$
& $>$128 & $>$128 & $>$128 & $>$128 \\

HY-P60324 &
IGCRCIQAG\Daa{a}GRLFVER\Daa{r}-NH$_2$
& $>$128 & $>$128 & $>$128 & $>$128 \\

HY-P60325 &
S\Daa{k}CRVLFFPFIFVRPHGSMFCCR-NH$_2$
& $>$128 & \textbf{16/32} &\textbf{64/64}& $>$128 \\

HY-P60322 &
KKRKHLRKAAGW\Daa{a}\Daa{y}KKVGLWAL-NH$_2$
& $>$128 & \textbf{8/8} & \textbf{16/32 }& $>$128 \\

HY-P60326 &
WFPHIARAKNGVRYA\Daa{k}NKKGKTVLK-NH$_2$
& $>$128 & $>$128 & $>$128 & $>$128 \\

Thanatin &
GSKKPVPIIYCNRRTGKCQRM-NH$_2$
& $>$128 & $>$128 & $>$128 & $>$128 \\

\bottomrule
\end{tabular}

\vspace{-5pt}
\end{table*}

\textbf{Results.} Five AMPGAN v3 candidates were synthesized and evaluated against clinically relevant bacterial strains. Minimum inhibitory and bactericidal concentrations (MIC/MBC) are reported in \cref{tab:mic_mbc}, alongside thanatin and conventional antibiotic controls. Two of five candidates exhibited antimicrobial activity. HY-P60322 showed potent activity against the Gram-positive strains \textit{B. subtilis} (MIC/MBC = 8/8~$\mu\text{g/mL}$) and \textit{S. epidermidis} (16/32~$\mu\text{g/mL}$). HY-P60325 showed activity against \textit{B. subtilis} (16/32~$\mu\text{g/mL}$) and \textit{S. epidermidis} (64/64~$\mu\text{g/mL}$). The remaining three candidates (HY-P60323, HY-P60324, HY-P60326) showed no activity ($>128$~$\mu\text{g/mL}$).

These results confirm that AMPGAN v3 generates sequences with genuine antimicrobial function, including across non-canonical chemistry (D-amino acids, C-terminal amidation) not supported by prior generative AMP models. However,  neither active candidate inhibited growth of the species used as its conditioning target: HY-P60322 (\textit{E. coli}-conditioned) was inactive against \textit{E. coli}, and HY-P60325 (\textit{S. aureus}-conditioned) was inactive against \textit{S. aureus}. Instead, both targeted Gram-positive species absent from their conditioning. Target-species conditioning therefore did not translate to species-selective activity. 

\subsection{\textit{In Silico} Evaluation}
\begin{figure}[h]
    \centering
    \includegraphics[width=1.0\columnwidth]{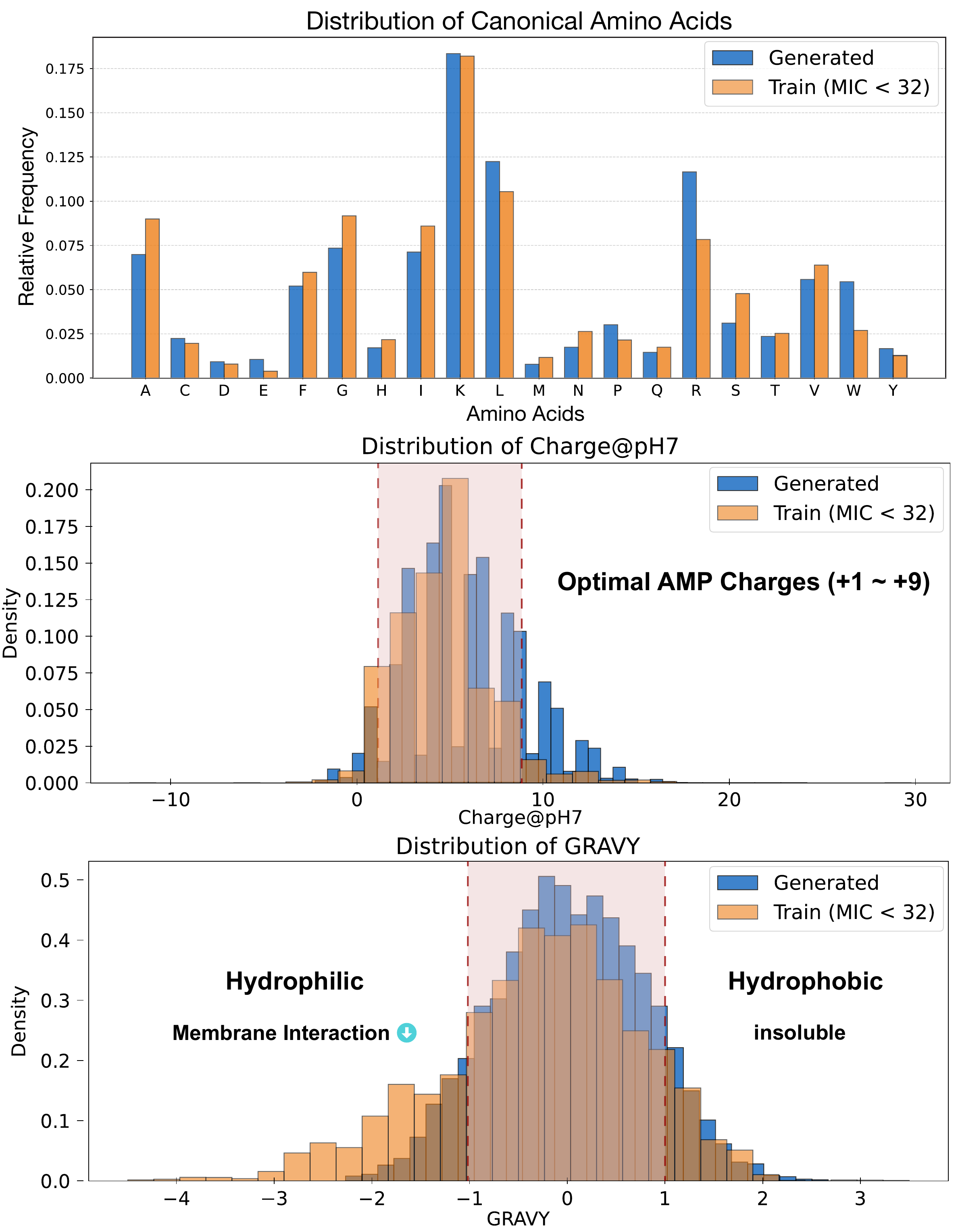}
    \caption{\textbf{Sequence property distributions for generated and training AMPs.} Per-residue frequencies (Top), net charge at pH~7 (center), and GRAVY hydrophobicity (Bottom), Generated sequences concentrate within AMP-active ranges charge in [+1, +9]; GRAVY in [-1, +1]).\textit{Top:} per-residue relative frequencies closely match the training distribution, except for Arg~(R) and Trp~(W), which are overrepresented by $\sim$ 0.025.}
    \label{fig:histo}
\end{figure}

\textbf{Generated Samples Quality Analysis:} \cref{fig:histo} compares the generated and training AMP distributions across three views: per-residue relative frequencies (top), net charge at pH~7 (center), and GRAVY hydrophobicity (bottom). AMPGAN v3 closely matches the training distribution on all three axes. At the residue level, the only notable deviations are Arginine~(R) and Tryptophan~(W), each overrepresented by approximately 0.025 (2.5 percentage points). At the sequence level, generated samples concentrate within the ranges associated with AMP activity, net charge in [+1,+9] and GRAVY in [-1,+1], preserving the cationic, amphipathic character required for selective interaction with negatively charged bacterial membranes.

\begin{table}[!t]
\caption{Training stability across $N=30$ random seeds. Both criteria measure post-training output quality: realistic, length-controllable generation. AMPGAN v2 numbers are reported in~\cite{ampganv2}.  }
\label{tab:stability}
\centering
\small
\begin{tabular}{llc}
\toprule
Model & Criterion & Successful runs \\
\midrule
AMPGAN v2 & entropy + length control & 3/30 \\
\midrule
\multirow{3}{*}{AMPGAN v3} 
 & \quad $p \geq 70\%$ & 21/30 \\
 & \quad $p \geq 80\%$ & 19/30 \\
 & \quad $p \geq 90\%$ & 13/30 \\
\bottomrule
\end{tabular}
\vspace{-5pt}

\end{table}
\textbf{Training Stability}: GAN training is known to suffer from mode collapse and instability~\cite{GAN_collapse}. We measure training stability across 30 random seeds for both AMPGAN v2 and AMPGAN v3 under matched experimental conditions, following the criterion of realistic sequence entropy and length-controllable generation defined in~\cite{ampganv2}. AMPGAN v2 succeeds in only 3/30 runs, consistent with the $\sim$10\% rate reported in the original work. Failed AMPGAN v2 runs typically collapse to repeating a single token across the entire output. AMPGAN v3 introduces structured tokens, which shift the dominant failure mode from global collapse to local formatting errors, such as incorrect placement of SOS, EOS, or terminus tokens. To quantify stability, we evaluate the proportion $p$ of generated sequences per run that satisfy the output validity criterion (i.e., containing a valid amino acid span with length error within 3 residues and properly structured tokens).

As shown in Table~\ref{tab:stability}, AMPGAN v3 achieves substantially higher stability across all thresholds. At $p \geq 70\%$, 21/30 runs are successful. Performance remains robust at stricter thresholds, with 19/30 runs at $p \geq 80\%$ and 13/30 runs at $p \geq 90\%$ (see Table~\ref{tab:stability} for AMPGAN v2 comparisons). These results indicate that AMPGAN v3 not only mitigates catastrophic mode collapse but also produces consistently valid outputs across the majority of training runs.

\begin{table}[ht]
\caption{Predicted AMP rates (\%) on $N=2{,}000$ unconditional generations from each model.}
\label{tab:external_predictor}
\centering
\small
\begin{tabular}{lcc}
\toprule
Model & AMP Scanner v2 & amPEP \\
\midrule
\textbf{AMPGAN v3} & \textbf{92.7} & \textbf{95.1} \\
\midrule
AMPGAN v2 & 80.0 & 83.4 \\
HydrAMP & 81.0 & 83.0 \\
AMP-Designer & 92.4 & 83.9 \\
\bottomrule
\end{tabular}

\end{table}
\textbf{External Predictor Scores}: To assess whether AMPGAN v3 generates AMP-like sequences, we classified generated outputs using two independent predictors that differ in both architecture and training data: AMP Scanner v2~\cite{ampscanner}, a Conv+LSTM deep learning classifier, and amPEP~\cite{ampep}, a random forest model based on amino-acid distribution patterns. For comparison, we generated sequences from AMPGAN v2~\cite{ampganv2} and HydrAMP~\cite{hydramp} under their default settings, and utilized the pre-generated sequences released with AMP-Designer~\cite{AMP_GPT}. For each model, we report the fraction of $N=2{,}000$ unconditional generations classified as AMP (\cref{tab:external_predictor}).

AMPGAN v3 achieves the highest predicted-AMP rate on both classifiers (92.7\% / 95.1\%), outperforming AMPGAN v2 (80.0\% / 83.4\%), HydrAMP (81.0\% / 83.0\%), and AMP-Designer (92.4\% / 83.9\%). Agreement across two architecturally distinct predictors suggests the gain is not an artifact of either classifier's inductive bias. We hypothesize that the multi-discriminator design contributes to this improvement by separating adversarial from activity-aware supervision.

\textbf{Embedding-space Distribution Analysis:} Beyond per-residue and physicochemical statistics, we evaluate whether AMPGAN v3 generations occupy the same region of sequence space as real AMPs. We embed real AMPs, random sequences, and generations from AMPGAN v2~\cite{ampganv2}, HydrAMP~\cite{hydramp}, AMP-Designer~\cite{AMP_GPT}, and AMPGAN v3 using ESM2~\cite{ESM2}.  These embeddings are then projected to 32 PCA components, and estimate Kullback-Leibler divergences (k=3) against the real distribution using a k-nearest neighbor density ratio estimator~\cite{knn}. Forward $KL(\text{real}|\text{gen})$ measures coverage (i.e., How well generations span the real distribution) while reverse $KL(\text{gen}|\text{real})$ measures fidelity (i.e., Whether generations remain on the real-AMP manifold).
\begin{table}[htbp]
\caption{KL divergence between generations and real AMPs in ESM2 embedding space ($k$=10, 32-d PCA, $n$=2000/group). Lower is better. $\Delta$ values are paired bootstrap differences (baseline $-$ v3) with 95\% CI over 1000 resamples; positive favors v3.}
\label{tab:divergence}
\centering
\footnotesize
\setlength{\tabcolsep}{3pt}
\begin{tabular}{lccc}
\toprule
 & Forward $\downarrow$ & Reverse $\downarrow$ \\
 & (coverage) & (fidelity) & \\
\midrule
v3 (Ours)      & 10.90 & 6.43 \\
\midrule
v2             & 11.08 & 7.56  \\
HydrAMP        & 14.94 & 8.82 \\
AMP-Designer   & \textbf{7.22}  & \textbf{2.75}  \\
Random         & 22.08 & 11.20\\
\midrule
$\Delta$ (v2$-$v3)           & $+0.18$ & $\mathbf{+1.13}$ \\
95\% CI                        & [$-0.42$, $+0.75$] & [$\mathbf{+0.72}$, $\mathbf{+1.56}$] \\
\midrule
$\Delta$ (HydrAMP$-$v3)      & $\mathbf{+4.04}$ & $\mathbf{+2.39}$ \\
95\% CI                        & [$\mathbf{+3.14}$, $\mathbf{+4.90}$] & [$\mathbf{+1.82}$, $\mathbf{+2.91}$]  \\
\midrule
$\Delta$ (AMP-Designer$-$v3) & $-3.68$ & $-3.68$  \\
95\% CI                        & [$-4.14$, $-3.22$] & [$-4.01$, $-3.20$] \\
\bottomrule
\end{tabular}
\vspace{-10pt}
\end{table}

All four generative models substantially outperform random sequences (symmetric KL 4.99-11.88 vs. 16.64). Compared to its predecessor, V3 significantly improves fidelity (reverse KL 6.43 vs 7.56, paired bootstrap 95\% CI on the difference $[+0.72, +1.56]$) while matching V2's coverage (forward KL difference 95\% CI $[-0.42, +0.75]$, not significant). V3 also outperforms HydrAMP on both axes (forward KL difference $[+3.14, +4.90]$, reverse KL $[+1.82, +2.91]$, both 95\% CIs excluding zero). AMP-Designer achieves the lowest divergence on both axes (reverse KL 2.75; paired difference vs v3 $[-4.01, -3.20]$), which we attribute to its initialization from AMP-GPT, a transformer pretrained on 600K+ UniProt peptides, approximately $63\times$ the size of our labeled training set, providing a strong protein prior that aligns closely with ESM2's pretraining distribution. AMPGAN v3, trained from scratch on $\sim$10K AMPs without protein prior pretraining, is structurally disadvantaged on this metric independent of antimicrobial properties.

\subsection{Agentic Workflow Evaluation}

\begin{figure*}[h]
    \centering
\includegraphics[width=1.0\textwidth]{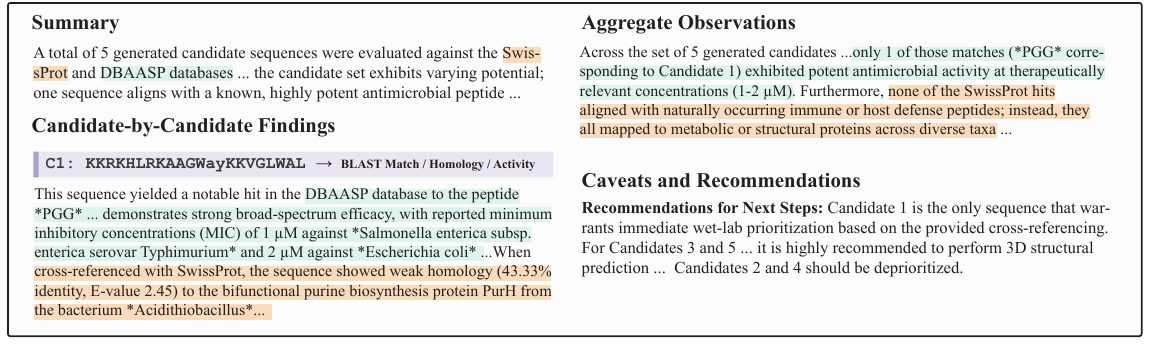}
\caption{\textbf{Multi-agent final report.} The Planning Agent synthesizes BLAST hits from the Verifying Agent into per-candidate findings, aggregate observations, and prioritization recommendations. Representative trajectory can be found in~\cref{app:agentic-trajectory}}
    \label{fig:report}
\vspace{-10pt}
\end{figure*}
We present preliminary observations on the agentic workflow, comparing our multi-agent system against two single-agent baselines that progressively combine planning and execution into one model, allowing us to examine which architectural choices affect efficiency. Our goal is to characterize how the pipeline behaves on a representative AMP discovery task and to illustrate the kind of output it produces, leaving prospective evaluation of agent-prioritized candidates to future work.

\textbf{Settings:} (1) Single-agent (batched): One LLM plans and executes inline, issuing multiple tool calls per iteration. Tests whether unified planning and execution in a single model is more efficient than decomposition. (2) Single-agent (stepwise): Same single LLM, but restricted to exactly one tool call per iteration. Isolates the effect of role decomposition by removing batched execution while retaining a unified model.
\textbf{(3) Multi-agent (Ours): }Planning and execution decomposed across separate models, with role-specific tool subsets per executor.

All settings receive the same objective: generate $K$ candidate AMP sequences with D-amino acids targeting \textit{E.coli} with $\alpha$-helical secondary structure. Single-agent baselines use \textit{Gemini 3.1 Pro Preview}~\cite{gemini_pro} throughout, since planning is required at every step. The multi-agent setting uses \textit{Gemini 3.1 Pro Preview} for the Planner and \textit{Gemini 3.1 Flash Lite Preview}~\cite{gemini_flash} for executors. All runs across all settings returned the requested $K$ candidates, so cost metrics are directly comparable. We exclude \texttt{Filter\_Status} calls from tool-call counts: it is a bookkeeping tool the agent invokes after each filtering request to track progress, but single-agent settings invoke it redundantly after each filter call, inflating their counts. Removing these calls ensures the comparison reflects meaningful tool calls rather than bookkeeping overhead.

\begin{figure}[h]
    \centering
    \includegraphics[width=\columnwidth]{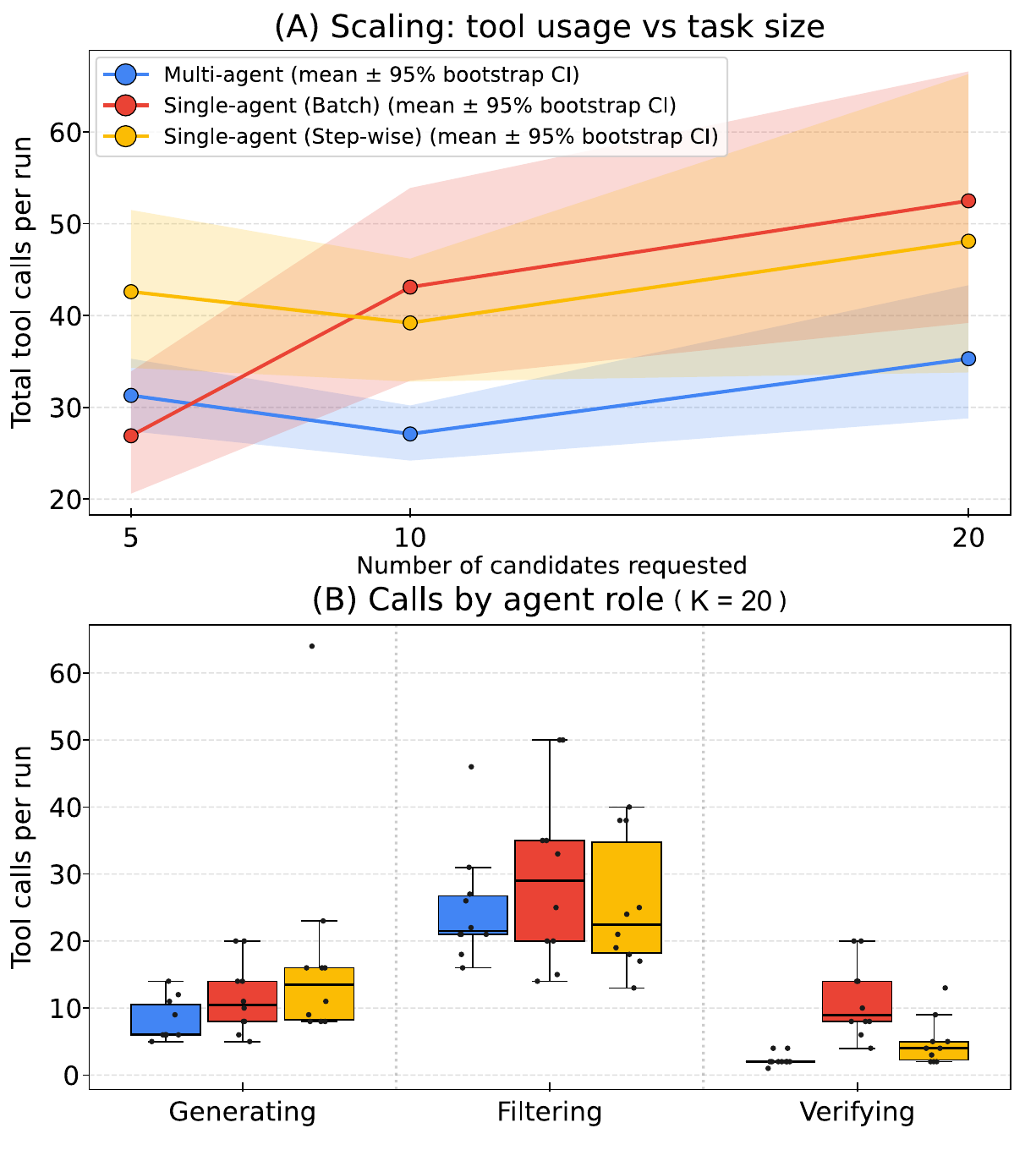}
    \caption{\textbf{Total tool calls per run as a function of task size.} Each setting was run across $N=10$ independent seeds at $K \in \{5, 10, 20\}$ requested candidates. Lines show per-seed mean tool calls; shaded bands show 95\% bootstrap confidence intervals; individual points show per-seed counts. }
    \label{fig:scaling}
    \vspace{-10pt}
\end{figure}

\textbf{Tool Call Efficiency: }We ran each setting across $N=10$ independent seeds at three task sizes $K= \{5, 10, 20\}$ requested candidates. Each tool call corresponds to one decision-execution cycle: The LLM selects a tool, the tool executes, and the result is returned to the agent's context for the next decision (see~\cref{sec:methodology_agentic} for more details). \cref{fig:scaling} (Top) shows total tool calls per run. Multi-agent uses fewer tool calls than batch across all task sizes. Both batch and stepwise grow and show substantially wider bootstrap intervals as K increases. Decomposing $K=20$ by agent role (\cref{fig:scaling}, Bottom) localizes the multi-agent advantage to the verification stage; filtering and generation are comparable across all three settings.

\begin{table}[h]
\centering
\small
\caption{Output token usage per run across orchestration settings ($K=20, N=10$ seeds). Multi-agent splits generation between a Planner (large model) and Executors (small model), enabling cost-effective routing. Single-agent settings must route all tokens through a large model.}
\label{tab:tokens}
\begin{tabular}{lrrr}
\toprule
\textbf{Setting} & \textbf{Mean} & \textbf{Median} & \textbf{Std}  \\
\midrule
\multicolumn{4}{l}{\textit{Multi-agent (Ours)}} \\
\quad Planner (large) & 3{,}724 & 3{,}226 & 1{,}298  \\
\quad Executors (small) & 12{,}585 & 11{,}044 & 3{,}037  \\
\midrule
\multicolumn{4}{l}{\textit{Single-agent (large model only)}} \\
\quad Batched & 9{,}942 & 9{,}400 & 3{,}579  \\
\quad Stepwise & 24{,}845 & 21{,}169 & 13{,}733  \\
\bottomrule
\end{tabular}

\end{table}

\textbf{Cost-effective model routing.} Tool-call counts alone understate the cost difference between settings. \cref{tab:tokens} reports output token usage per run at $K=20$. In the multi-agent setting, the Planner produces only $\sim$3.7K tokens per run while Executors produce $\sim$12.6K. However, Executors merely structure tool calls following the Planner's instructions, so they can be backed by a smaller, cheaper model. Only the Planner, which performs cross-step reasoning, requires a large model. The single-agent baselines have no such separation: batched produces $\sim$9.9K and stepwise $\sim$24.8K tokens, all of which must route through the large model. This model flexibility, large model for planning and small model for execution, is the architectural advantage that distinguishes multi-agent from single-agent in practice.

\textbf{Qualitative Results}: Beyond orchestration cost, we examine the content of the multi-agent workflow's output. \cref{fig:report} shows a representative final report generated by running the Verification pipeline on the five \textit{in vitro} candidates from \cref{sec:invitro}. The Verifying Agent retrieves BLAST~\cite{blast} hits from SwissProt~\cite{Swissprot} and DBAASP~\cite{dbaasp} for each candidate, and the Planning Agent synthesizes these into per-candidate findings, aggregate observations, and prioritization recommendations. For HY-P60322, the report highlights a DBAASP hit to PGG (reported MIC 1–2 $~\mu$M against S. enterica and E. coli) and recommends wet-lab prioritization. HY-P60325 is deprioritized after mapping to a sequence with no reported antimicrobial activity. The report's prioritization is consistent with our \textit{in vitro} results: HY-P60322 was one of the two active candidates, while three of the four candidates flagged for deprioritization or further analysis showed no activity.
This result qualitatively supports the effectiveness of PepCraft. Prospective evaluation of selection ability remains a task for future research.


\section{Conclusion and Future Work}

In this study, we propose AMPGAN v3, which generates antimicrobial peptides spanning canonical and non-canonical chemistry, including D-amino acids and N/C-terminus modifications. Beyond expanding the chemical space, AMPGAN v3 improves training stability from 10\% to 70\% (\cref{tab:stability}) of its predecessor and outperforms competing models in producing AMP-like sequences (\cref{tab:external_predictor}). Wet-lab evaluation of five candidates yields two actives, notably HY-P60322 with an MIC of 8~$\mu$g/mL against \textit{B. subtilis} (\cref{tab:mic_mbc}), though neither active candidate inhibited its conditioning target species. While our proposed PepCraft framework is preliminary, the qualitative consistency between the agent's prioritization and these \textit{in vitro} outcomes is encouraging (\cref{fig:report}). We anticipate PepCraft will grow as the ecosystem expands to include automated model training, molecular dynamics simulations, and broader peptide design tasks.

Future work will prioritize wet-lab validation of agent-selected candidates and the refinement of target-aware peptide generation. We will also expand the chemical vocabulary to include post-translational modifications, such as cyclization, to better support therapeutic applications.

\section{Acknowledgments}
This work was supported by NIH R01 award~(R01GM143370). S.T.S. received partial support from an NIH R35 Award (GM147579). The agentic work flow computation was supported by Google Cloud Research Credits program with the award GCP19980904.
\newpage

\bibliography{references}

@article{dbaasp, title={DBAASP v3: database of antimicrobial/cytotoxic activity and structure of peptides as a resource for development of new therapeutics}, volume={49}, url={https://academic.oup.com/nar/article/49/D1/D288/5957160}, DOI={https://doi.org/10.1093/nar/gkaa991}, number={D1}, journal={Nucleic Acids Research}, author={Pirtskhalava, Malak and Amstrong, Anthony A and Grigolava, Maia and Chubinidze, Mindia and Alimbarashvili, Evgenia and Vishnepolsky, Boris and Gabrielian, Andrei and Rosenthal, Alex and Hurt, Darrell E and Tartakovsky, Michael}, year={2020}, month={Nov}, pages={D288–D297} }

@article{ampganv2, title={AMPGAN v2: Machine Learning-Guided Design of Antimicrobial Peptides}, volume={61}, DOI={https://doi.org/10.1021/acs.jcim.0c01441}, number={5}, journal={Journal of Chemical Information and Modeling}, author={Van Oort, Colin M. and Ferrell, Jonathon B. and Remington, Jacob M. and Wshah, Safwan and Li, Jianing}, year={2021}, month={Mar}, pages={2198–2207} }

@article{ampgan_bio, title={A Generative Approach toward Precision Antimicrobial Peptide Design}, DOI={https://doi.org/10.1101/2020.10.02.324087}, journal={bioRxiv (Cold Spring Harbor Laboratory)}, publisher={Cold Spring Harbor Laboratory}, author={Ferrell, Jonathon B and Remington, Jacob M and Van, Colin M and Sharafi, Mona and Reem Aboushousha and Janssen-Heininger, Yvonne and Schneebeli, Severin T and Wargo, Matthew J and Safwan Wshah and Li, Jianing}, year={2020}, month={Oct} }

@article{antibiotics_investment, title={Current economic and regulatory challenges in developing antibiotics for Gram-negative bacteria}, volume={3}, DOI={https://doi.org/10.1038/s44259-025-00123-1}, number={1}, journal={npj Antimicrobials and Resistance}, publisher={Springer Science and Business Media LLC}, author={Gargate, Nupur and Laws, Mark and Rahman, Khondaker Miraz}, year={2025}, month={Jun} }

@article{discovery_void, title={Challenges of antibacterial discovery}, volume={24}, DOI={https://doi.org/10.1128/CMR.00030-10}, number={1}, journal={Clinical microbiology reviews}, publisher={American Society for Microbiology (ASM)}, author={Silver, Lynn L}, year={2011}, pages={71–109} }

@misc{WHO, title={Antimicrobial resistance}, url={https://www.who.int/news-room/fact-sheets/detail/antimicrobial-resistance}, author={WHO}, year={2023}, month={Nov} }

@article{AMR_kills_millions, title={Global Burden of Bacterial Antimicrobial Resistance in 2019: a Systematic Analysis}, volume={399}, DOI={https://doi.org/10.1016/S0140-6736(21)02724-0}, number={10325}, journal={The Lancet},   
 author  = {Christopher J. L. Murray and Kevin Shunji Ikuta and 
             Fablina Sharara and Lucien Swetschinski and 
             Gisela Robles Aguilar and others}, year={2022}, month={Feb}, pages={629–655} }

@article{AMP_membranes, title={Antimicrobial Peptides: Opportunities and Challenges in Overcoming Resistance}, volume={286}, DOI={https://doi.org/10.1016/j.micres.2024.127822}, journal={Microbiological Research}, author={Bucataru, Cezara and Ciobanasu, Corina}, year={2024}, month={Jun}, pages={127822} }

@article{hydramp, title={Discovering highly potent antimicrobial peptides with deep generative model HydrAMP}, volume={14}, DOI={https://doi.org/10.1038/s41467-023-36994-z}, number={1}, journal={Nature Communications}, author={Szymczak, Paulina and Możejko, Marcin and Grzegorzek, Tomasz and Jurczak, Radosław and Bauer, Marta and Neubauer, Damian and Sikora, Karol and Michalski, Michał and Sroka, Jacek and Setny, Piotr and Kamysz, Wojciech and Szczurek, Ewa}, year={2023}, month={Mar} }

@article{ESM2, title={Evolutionary-scale prediction of atomic-level protein structure with a language model}, volume={379}, DOI={https://doi.org/10.1126/science.ade2574}, number={6637}, journal={Science}, author={Lin, Zeming and Akin, Halil and Rao, Roshan and Hie, Brian and Zhu, Zhongkai and Lu, Wenting and Smetanin, Nikita and Verkuil, Robert and Kabeli, Ori and Shmueli, Yaniv and dos Santos Costa, Allan and Fazel-Zarandi, Maryam and Sercu, Tom and Candido, Salvatore and Rives, Alexander}, year={2023}, month={Mar}, pages={1123–1130} }

@article{AMP_GPT, title={Discovery of antimicrobial peptides with notable antibacterial potency by an LLM-based foundation model}, volume={11}, DOI={https://doi.org/10.1126/sciadv.ads8932}, number={10}, journal={Science Advances}, author={Wang, Jike and Feng, Jianwen and Kang, Yu and Pan, Peichen and Ge, Jingxuan and Wang, Yan and Wang, Mingyang and Wu, Zhenxing and Zhang, Xingcai and Yu, Jiameng and Zhang, Xujun and Wang, Tianyue and Wen, Lirong and Yan, Guangning and Deng, Yafeng and Shi, Hui and Hsieh, Chang-Yu and Jiang, Zhihui and Hou, Tingjun}, year={2025}, month={Mar} }

@article{moformer, title={MOFormer: navigating the antimicrobial peptide design space with Pareto-based multi-objective transformer}, volume={26}, DOI={https://doi.org/10.1093/bib/bbaf376}, number={6}, journal={Briefings in Bioinformatics}, author={Wang, Li and Fu, Xiangzheng and Yang, Jiahao and Zhang, Xinyi and Ye, Xiucai and Sakurai, Tetsuya and Zeng, Xiangxiang and Liu, Yiping}, year={2025}, month={Nov} }

@article{autolab,
      title={AutoLabs: Cognitive Multi-Agent Systems with Self-Correction for Autonomous Chemical Experimentation}, 
      author={Gihan Panapitiya and Emily Saldanha and Heather Job and Olivia Hess},
      year={2025},
      eprint={2509.25651},
      archivePrefix={arXiv},
      primaryClass={cs.AI},
      doi={ 	
https://doi.org/10.48550/arXiv.2509.25651}, 
journal={arXiv}
}

@INPROCEEDINGS{GAN_collapse,
  author={Kossale, Youssef and Airaj, Mohammed and Darouichi, Aziz},
  booktitle={2022 8th International Conference on Optimization and Applications (ICOA)}, 
  title={Mode Collapse in Generative Adversarial Networks: An Overview}, 
  year={2022},
  volume={},
  number={},
  pages={1-6},
  doi={10.1109/ICOA55659.2022.9934291}}

@article {bioreason_pro,
	author = {Fallahpour, Adibvafa and Seyed-Ahmadi, Arman and Idehpour, Parsa and Ibrahim, Omar and Gupta and others},
	title = {BioReason-Pro: Advancing Protein Function Prediction with Multimodal Biological Reasoning},
	elocation-id = {2026.03.19.712954},
	year = {2026},
	doi = {10.64898/2026.03.19.712954},
	publisher = {Cold Spring Harbor Laboratory},
	eprint = {https://www.biorxiv.org/content/early/2026/03/20/2026.03.19.712954.full.pdf},
	journal = {bioRxiv}
}

@article{AMP_fact,
    author = {Wang, Guangshun and Schmidt, Cindy and Li, Xia and Wang, Zhe},
    title = {APD6: the antimicrobial peptide database is expanded to promote research and development by deploying an unprecedented information pipeline},
    journal = {Nucleic Acids Research},
    volume = {54},
    number = {D1},
    pages = {D363-D374},
    year = {2026},
    month = {01},
    issn = {1362-4962},
    doi = {10.1093/nar/gkaf860},
    url = {https://doi.org/10.1093/nar/gkaf860},
    eprint = {https://academic.oup.com/nar/article-pdf/54/D1/D363/64235225/gkaf860.pdf},
}

@article{AMP_chemical_fact, title={A Review of Antimicrobial Peptides: Structure, Mechanism of Action, and Molecular Optimization Strategies}, volume={10}, DOI={https://doi.org/10.3390/fermentation10110540}, number={11}, journal={Fermentation}, publisher={Multidisciplinary Digital Publishing Institute}, author={Ma, Xu and Wang, Qiang and Ren, Kexin and Xu, Tongtong and Zhang, Zigang and Xu, Meijuan and Rao, Zhiming and Zhang, Xian}, year={2024}, month={Oct}, pages={540–540} }

@article{AMP_fast, title={Kinetics of antimicrobial peptide activity measured on individual bacterial cells using high-speed atomic force microscopy}, volume={5}, DOI={https://doi.org/10.1038/nnano.2010.29}, number={4}, journal={Nature nanotechnology}, publisher={Nature Portfolio}, author={Fantner, Georg E and Barbero, Roberto J and Gray, David S and Belcher, Angela M}, year={2010}, month={Mar}, pages={280–285} }

@article{pepgan,
author = {Tucs, Andrejs and Tran, Duy Phuoc and Yumoto, Akiko and Ito, Yoshihiro and Uzawa, Takanori and Tsuda, Koji},
title = {Generating Ampicillin-Level Antimicrobial Peptides with Activity-Aware Generative Adversarial Networks},
journal = {ACS Omega},
volume = {5},
number = {36},
pages = {22847-22851},
year = {2020},
doi = {10.1021/acsomega.0c02088},
}

@article{HMAMP,
author = {Wang, Li and Liu, Yiping and Fu, Xiangzheng and Ye, Xiucai and Shi, Junfeng and Yen, Gary G. and Zou, Quan and Zeng, Xiangxiang and Cao, Dongsheng},
title = {HMAMP: Designing Highly Potent Antimicrobial Peptides Using a Hypervolume-Driven Multiobjective Deep Generative Model},
journal = {Journal of Medicinal Chemistry},
volume = {68},
number = {8},
pages = {8346-8360},
year = {2025},
doi = {10.1021/acs.jmedchem.4c03073},
    note ={PMID: 40232176},
}

@InProceedings{film,
  title={FiLM: Visual Reasoning with a General Conditioning Layer},
  author={Ethan Perez and Florian Strub and Harm de Vries and Vincent Dumoulin and Aaron C. Courville},
  booktitle={AAAI},
  year={2018}
}

@inproceedings{
proteincrow,
title={ProteinCrow: A Language Model Agent That Can Design Proteins},
author={Manvitha Ponnapati and Sam Cox and Cade W Gordon and Michael J Hammerling and Siddharth Narayanan and Jon M Laurent and James D. Braza and Michaela M. Hinks and Michael D Skarlinski and Samuel G Rodriques and Andrew White},
booktitle={ICML 2025 Generative AI and Biology (GenBio) Workshop},
year={2025},
url={https://openreview.net/forum?id=ljXgWDtqCu}
}

@article{ampscanner, title={Deep learning improves antimicrobial peptide recognition}, volume={34}, DOI={https://doi.org/10.1093/bioinformatics/bty179}, number={16}, journal={Bioinformatics}, author={Veltri, Daniel and Kamath, Uday and Shehu, Amarda}, editor={Hancock, John}, year={2018}, month={Mar}, pages={2740–2747} }

@article{ampep, title={AmPEP: Sequence-based prediction of antimicrobial peptides using distribution patterns of amino acid properties and random forest}, volume={8}, DOI={https://doi.org/10.1038/s41598-018-19752-w}, number={1}, journal={Scientific Reports}, author={Bhadra, Pratiti and Yan, Jielu and Li, Jinyan and Fong, Simon and Siu, Shirley W. I.}, year={2018}, month={Jan} }

@article{alphafold3, title={Accurate structure prediction of biomolecular interactions with AlphaFold 3}, volume={630}, DOI={https://doi.org/10.1038/s41586-024-07487-w}, number={630}, journal={Nature}, author={Abramson, Josh and Adler, Jonas and Dunger, Jack and Evans, Richard and Green, Tim and Pritzel, Alexander and others}, year={2024}, month={May}, pages={493–500} }

@inproceedings{simplefold,
title = {SimpleFold: Folding Proteins is Simpler than You Think},
booktitle = {ICLR},
author = {Yuyang Wang and Jiarui Lu and Navdeep Jaitly and Josh Susskind and Miguel Angel Bautista},
year = {2025},
URL = {https://arxiv.org/abs/2509.18480v1}
}

@Article{peptide_need_modifications,
AUTHOR = {Lucana, Maria C. and Arruga, Yolanda and Petrachi, Emilia and Roig, Albert and Lucchi, Roberta and Oller-Salvia, Benjamí},
TITLE = {Protease-Resistant Peptides for Targeting and Intracellular Delivery of Therapeutics},
JOURNAL = {Pharmaceutics},
VOLUME = {13},
YEAR = {2021},
NUMBER = {12},
ARTICLE-NUMBER = {2065},
URL = {https://www.mdpi.com/1999-4923/13/12/2065},
PubMedID = {34959346},
ISSN = {1999-4923},

DOI = {10.3390/pharmaceutics13122065}
}

@article{alpha_helical_instable, title={Design of Stable $\alpha$-Helical Peptides and Thermostable Proteins in Biotechnology and Biomedicine}, volume={8}, DOI={https://doi.org/10.32607/20758251-2016-8-4-70-81}, number={4}, journal={Acta Naturae}, author={Yakimov, A.P. and Afanaseva, A.S. and Khodorkovskiy, M.A. and Petukhov, M.G.}, year={2016}, month={Dec}, pages={70–81} }

@article{Swissprot, title={UniProtKB/Swiss-Prot}, volume={406}, url={https://europepmc.org/article/MED/18287689}, DOI={https://doi.org/10.1007/978-1-59745-535-0_4}, journal={Methods in molecular biology (Clifton, N.J.)}, author={Boutet, Emmanuel and Lieberherr, Damien and Tognolli, Michael and Schneider, Michel and Bairoch, Amos}, year={2007}, month={Jan}, pages={89–112} }

@article{blast, title={BLAST+: architecture and applications}, volume={10}, number={1}, journal={BMC Bioinformatics}, author={Camacho, Christiam and Coulouris, George and Avagyan, Vahram and Ma, Ning and Papadopoulos, Jason and Bealer, Kevin and Madden, Thomas L}, year={2009}, pages={421} }

@article{knn, title={Introduction to Machine learning: k-nearest Neighbors}, volume={4}, DOI={https://doi.org/10.21037/atm.2016.03.37}, number={11}, journal={Annals of Translational Medicine}, author={Zhang, Zhongheng}, year={2016}, month={Jun}, pages={218–218} }

@misc{gemini_pro,
  author       = {{Google Cloud}},
  title        = {Gemini 3.1 Pro},
  howpublished = {\url{https://docs.cloud.google.com/vertex-ai/generative-ai/docs/models/gemini/3-1-pro}},
  year         = {2026},
  note         = {Accessed: 2026-05-05}
}

@misc{gemini_flash,
  author       = {{Google Cloud}},
  title        = {Gemini 3.1 Flash},
  howpublished = {\url{https://docs.cloud.google.com/vertex-ai/generative-ai/docs/models/gemini/3-1-flash-lite}},
  year         = {2026},
  note         = {Accessed: 2026-05-05}
}

@manual{maestro,
  title        = {Maestro, {S}chrödinger, LLC Release 2026-2},
  howpublished = {\url{https://www.schrodinger.com/platform/products/maestro/}},
  author       = {{Schrödinger, LLC}},
  address      = {New York, NY},
  year         = {2025},
  note         = {Accessed: 2026-03-01}
}
\bibliographystyle{icml2026}

\newpage
\appendix
\onecolumn  

\section{In Vitro Experimental Protocol}

\label{app:invitro}
Selected peptide candidates were commercially synthesized by a MedChemExpress(MCE) using standard solid-phase peptide synthesis and purified by high-performance liquid chromatography (HPLC). Peptide identity and purity were confirmed by mass spectrometry and analytical HPLC according to the quality-control reports.

\begin{table}[h]
\centering
\caption{Bacterial isolates used in the current study.}
\label{tab:strains}
\begin{tabular}{llll}
\toprule
Bacteria & Strain ID & Strain Designation & Source \\
\midrule
\textit{S. aureus}  & USA300-0114 & NRD384, MRSA & Wound, Mississippi, USA \\
\textit{B. subtilis}      & --- & --- & --- \\
\textit{S. epidermidis}   & ATCC 35984 & NRS 101 & Catheter sepsis, Tennessee, USA \\
\textit{E. faecium}       & HM-952 & 503 & Human, USA \\
\textit{E. coli}           & ATCC 25922   & Seattle 1946    & Washington, USA \\
\bottomrule
\end{tabular}
\end{table}
\vspace{-10pt}

\section{AMPGAN v3: Implementation Details and Additional Analyses}
All training was conducted on an NVIDIA RTX PRO 6000 Blackwell GPU, 
and all inference was performed on an NVIDIA GeForce RTX 4080.

\label{app:ampgan}

\subsection{Note on D- vs L-Amino Acids in Evaluation}
\label{app:ampgan-daa-note}

A central contribution of AMPGAN v3 is the ability to generate sequences containing
non-canonical residues (D-amino acids) and N/C-terminus modifications. We emphasize a
distinction between two settings in this paper:
\vspace{-2pt}
\begin{itemize}
\item \textbf{In vitro evaluation (Section~5.1).} Candidates synthesized for wet-lab
testing were drawn from the \emph{full} generative vocabulary, including D-amino
acid substitutions and C-terminal amidation, since these modifications are
indispensable for therapeutic viability \citep{peptide_need_modifications}. This setting reflects
the intended use case of AMPGAN v3.

\item \textbf{In silico evaluation (Section~5.2 and Tables~4--5).} For all
quantitative comparisons against AMPGAN v2, HydrAMP, and AMP-Designer, we generate
sequences using the \emph{L-amino acid subset} of our vocabulary only.
None of the baseline models support D-amino acids or terminal modifications, and
the external classifiers (AMP Scanner v2, amPEP) and embedding models (ESM2) used
for evaluation are trained exclusively on canonical L-peptides. Including
non-canonical residues in the comparison would have introduced an out-of-distribution
artifact that confounds rather than informs the comparison. Restricting to L-only
generations ensures the gains we report reflect modeling improvements rather than
vocabulary differences.
\end{itemize}
\vspace{-2pt}
\subsection{Hyperparameters and Training Configuration}

\begin{table}[h]
\centering
\caption{AMPGAN v3 training hyperparameters.}
\label{tab:hparams}
\begin{tabular}{ll@{\hskip 3em}ll}
\toprule
\multicolumn{2}{l}{\textbf{Generator}} & \multicolumn{2}{l}{\textbf{Discriminators}} \\
\midrule
Latent dim $d_z$        & 256    & $d_\text{model}$       & 64 \\
Condition embed $d_e$   & 128    & FFN hidden $d_\text{hid}$ & 256 \\
Transformer L / H       & 4 / 4  & Transformer L / H      & 4 / 4 \\
Hidden dim              & 128    & MLP head dim           & 64 \\
Gumbel temp $\tau$      & 1.0    & Dropout                & 0.2 \\
\midrule
\multicolumn{2}{l}{\textbf{Loss weights}} & \multicolumn{2}{l}{\textbf{Optimization}} \\
\midrule
$\lambda_\ell$ (length)   & 0.5  & Optimizer & AdamW, $\beta{=}(0.5, 0.999)$, wd $10^{-4}$ \\
$\lambda_s$ (similarity)  & 0.5  & LR (G / $D_\text{adv}$ / $D_\text{mic}$) & $10^{-4}$ / $5{\times}10^{-6}$ / $10^{-4}$ \\
$\lambda_a$ (adversarial) & 2    & Batch size      & 256 \\
$\lambda_m$ (MIC)         & 1    & Epochs          & 200 \\
$\alpha$ (padding)        & 0.1  & D : G updates   & 1 : 1 \\
\bottomrule
\end{tabular}
\end{table}
\label{app:ampgan-hparam}

\subsection{Token Vocabulary}
\label{app:ampgan-vocab}

Each sequence is encoded as
\texttt{[N-term][SOS]\,[Amino Acid Tokens]\,[EOS][C-term]}, with maximum length $L = 64$
residues. The full vocabulary (Table~\ref{tab:vocab}) contains $V = $ 48 tokens
spanning canonical L-amino acids, D-amino acid stereoisomers, sequence-boundary
markers, padding, and N/C-terminus chemical modifications. Lowercase letters denote
D-amino acids, preserving stereochemistry directly in the token stream. This unified
representation allows AMPGAN v3 to jointly generate sequence content and chemical
modifications within a single output stream, in contrast to prior generative AMP
models that operate over the 20 canonical L-amino acids only.
 
\begin{table}[h]
\centering
\caption{AMPGAN v3 token vocabulary. Lowercase tokens denote D-amino acid
stereoisomers of the corresponding uppercase L-amino acid. \texttt{<C8>}--\texttt{<C16>}
denote fatty-acid acylation at the N-terminus with the indicated chain length.}
\label{tab:vocab}
\begin{tabular}{ll}
\toprule
Category & Tokens \\
\midrule
L-amino acids (20)
  & \texttt{A, C, D, E, F, G, H, I, K, L, M, N, P, Q, R, S, T, V, W, Y} \\
D-amino acids (16)
  & \texttt{a, c, f, h, i, k, l, n, p, q, r, s, t, v, w, y} \\
Sequence boundaries
  & \texttt{<SOS>}, \texttt{<EOS>} \\
Sequence padding
  & \texttt{<blank>} \\
N-terminus modifications
  & \texttt{<ACT>} (acetylation), \texttt{<C8>}, \texttt{<C10>}, \texttt{<C12>},
    \texttt{<C14>}, \texttt{<C16>}, \texttt{<nblank>} \\
C-terminus modifications
  & \texttt{<AMD>} (amidation), \texttt{<cblank>} \\
\bottomrule
\end{tabular}
\end{table}

\label{app:ampgan-esm}

\vspace{-10pt}
\section{Agentic Pipeline: Architecture, Prompts, and Trajectories}
\label{app:agentic}

\subsection{System Prompts}
\label{app:agentic-prompts}

\paragraph{Planning Agent.}
\begin{quote}\small\ttfamily
You are a **Planning Agent** orchestrating a multi-agent system for **Antimicrobial Peptide (AMP) Discovery**.

Your role is to coordinate the workflow by selecting the next agent and issuing structured instructions in strict JSON format.

\#\#\# Available Agents

\{Agent Description\}

\{Agent Context\}
---

\#\#\# Planning Logic
- If no peptide sequences exist, you MUST generate sequences first.
- If peptide sequences exist but are unfiltered, you MUST filter.
- If filtered results satisfy all constraints, you MUST select **END**.
- You MAY iterate between Agents to refine results.

---

\#\#\# Output Rules (STRICT)
- Output ONLY JSON
- No explanations, no extra text
- Must follow schema exactly
- Do NOT assume hidden state — rely only on explicitly provided input context

---

\#\#\# JSON Format

\{ \\
\hspace*{1em}"Planning": \{ \\
\hspace*{2em}"Agent": "Generating | Filtering | Verifying | END", \\
\hspace*{2em}"Purpose": "2--3 sentence description of the current objective for the selected agent.", \\
\hspace*{2em}"Context": \{ \\
\hspace*{3em}"instructions": "Provide clear, structured instructions for the selected agent to complete the assigned task. This should describe what to do, not raw input data.", \\
\hspace*{3em}"details": "Contains all execution-critical information such as user-provided inputs, parameters, constraints, tool-call arguments, or any supplemental data required for the agent to perform its function accurately." \\
\hspace*{2em}\} \\
\hspace*{1em}\} \\
\}

\end{quote}
\newpage
\paragraph{Execution Agent.}
\begin{quote}\small\ttfamily
You are a **\{Agent\} Agent** designed to execute tool-based biological workflows for **Antimicrobial Peptide (AMP)** tasks.
Your role is to follow the instructions provided by the **Planning Agent** (your supervisor) and perform concrete actions using appropriate biological tools.

\#\#\# Context
- **Planning Agent Instruction:** 

\{Instruction\}

- **Your Capabilities:** 

\{Agent Description\}

\#\#\# Rules
1. Output ONLY valid JSON matching the schema below — no prose, no explanation.
2. Each step is exactly one tool invocation. Steps are numbered sequentially from 1.
3. Use only tools listed in your Agent Description. All required parameters must be provided.
4. You may call the same tool multiple times if needed.
5. You may add calls to reporting/support tools from your description if they improve output quality, but never invent tools.

\#\#\# Output Schema

\{ \\
\hspace*{1em}"Agent": "\{Agent\}", \\
\hspace*{1em}"Objective": "<concise task summary>", \\
\hspace*{1em}"Steps": [ \\
\hspace*{2em}\{ \\
\hspace*{3em}"id": 1, \\
\hspace*{3em}"Tool": "<tool\_name>", \\
\hspace*{3em}"Input": \{ "<param>": "<value>" \} \\
\hspace*{2em}\} \\
\hspace*{1em}] \\
\}
\end{quote}

\subsection{Example Outputs}
\paragraph{Planning Agent.}

\begin{quote}
\ttfamily
\{ \\
\hspace*{1em}"Planning": \{ \\
\hspace*{2em}"Agent": "Generating", \\
\hspace*{2em}"Purpose": "Generate the initial batch of Antimicrobial Peptide sequences targeting E. coli using AMPGAN-v3.", \\
\hspace*{2em}"Context": \{ \\
\hspace*{3em}"instructions": "Use the AMPGAN\_v3 tool to generate an initial batch of sequences targeting E. coli. Generate exactly 5 sequences for the first pass, with lengths between 10 and 20 amino acids.", \\
\hspace*{3em}"details": \{ \\
\hspace*{4em}"min\_length": 10, \\
\hspace*{4em}"max\_length": 20, \\
\hspace*{4em}"num\_generations": 5, \\
\hspace*{4em}"species\_of\_interest": "ecoli", \\
\hspace*{4em}"folder\_path": "PATH" \\
\hspace*{3em}\} \\
\hspace*{2em}\} \\
\hspace*{1em}\} \\
\}
\end{quote}
\newpage
\paragraph{Execution Agent.}

\begin{quote}
\ttfamily
\{ \\
\hspace*{1em}"Agent": "Verifying", \\
\hspace*{1em}"Objective": "Cross-reference filtered AMP candidates against SwissProt and DBAASP databases to assess novelty and retrieve biological annotations.", \\
\hspace*{1em}"Steps": [ \\
\hspace*{2em}\{ \\
\hspace*{3em}"id": 1, \\
\hspace*{3em}"Tool": "Verify\_SwissProt", \\
\hspace*{3em}"Input": \{ \\
\hspace*{4em}"folder\_path": "PATH" \\
\hspace*{3em}\} \\
\hspace*{2em}\}, \\
\hspace*{2em}\{ \\
\hspace*{3em}"id": 2, \\
\hspace*{3em}"Tool": "Verify\_DBAASP", \\
\hspace*{3em}"Input": \{ \\
\hspace*{4em}"folder\_path": "PATH" \\
\hspace*{3em}\} \\
\hspace*{2em}\} \\
\hspace*{1em}] \\
\}
\end{quote}

\subsection{Executor Report Example}
\label{app:agentic_report}
\begin{quote}
\ttfamily
**Execution Overview**  
The objective of this workflow was to refine a set of generated antimicrobial peptides (AMPs) by applying a sequence of four specific filters: length, cationicity, hydrophobicity, and D-amino acid composition. These filters were applied to a dataset located at "PATH" to identify high-quality candidates.

**Step-wise Results**  
The length filter was applied to select sequences between 10 and 20 amino acids, followed by a cationicity filter (2.0 - 8.0) and a hydrophobicity filter (-0.5 - 0.5). Finally, a D-amino acid filter was executed. For each of these four steps, the original CSV file was updated with corresponding indicator columns ('length\_filter', 'cationicity\_filter', 'hydrophobicity\_filter', and 'damino\_filter'), where a value of 1 marks a successful pass and 0 denotes a failure to meet the criteria.

**Final Outcome**  
The final status assessment confirms that, from the initial pool of 5 generated sequences, 1 sequence successfully passed all applied filters. This sequence met the requirements for length, cationicity, hydrophobicity, and D-amino acid composition, representing the single candidate remaining after the full filtering process.

**Completed Tasks**  
The successfully completed tasks include the Length Filter, Cationicity Filter, Hydrophobicity Filter, D-amino acid Filter, and the final Filter Status report.
\end{quote}

\subsection{Tool Specifications}
\label{app:agentic-tools}

Each executor agent operates over a fixed set of tools registered in its system prompt. Tools accept structured JSON inputs from the agent and output string-template reports describing their outcome (e.g., counts of surviving candidates, BLAST hits, generation summaries). The executor agent aggregates these per-tool reports into a single run report, which is routed back to the Planning Agent as context for its next decision. This separation lets tools remain stateless and replaceable while the executor handles report composition. Below we list each agent's tools, their inputs, and operational notes that the Planning Agent uses when deciding whether to re-invoke.

\paragraph{Generating Agent.}
Translates user-defined natural-language constraints into structured tool inputs.

\begin{itemize}
\item \texttt{AMPGAN\_v3} — GAN-based AMP generator 
(\cref{sec:methodology}). Outputs are written to 
\texttt{generated\_sequences.csv} for downstream tools. 
\textit{Inputs}: \texttt{min\_length}, \texttt{max\_length} (int), 
\texttt{num\_generations} (int), 
\texttt{species\_of\_interest} (str; one of 
\texttt{ecoli}, \texttt{paeruginosa}, \texttt{kpneumoniae}, 
\texttt{saureus}, \texttt{bsubtilis}, \texttt{sepidermidis}), 
\texttt{folder\_path}.

\item \texttt{Predict\_Structure} — 3D structure prediction via SimpleFold~\cite{simplefold} 360M. Used only on filtered subsets to bound cost; the Planning Agent is instructed to defer this tool until after physicochemical filtering. \textit{Inputs}: \texttt{folder\_path}. \end{itemize}

The Generating Agent follows an adaptive batch-sizing policy: the first invocation generates the user-requested count $N$; after observing the post-filter pass rate $p$, subsequent rounds generate $\lceil S / p \rceil$ candidates to cover the shortfall $S$, capped at $3N$ per call.

\paragraph{Filtering Agent.}Applies sequential physicochemical and structural screens. Each filter writes a boolean column to the candidate CSV and the Filter\_Status tool reports the surviving count. The Planning Agent prioritizes cheap property filters before invoking structure-dependent filters.

\begin{itemize}
\item \texttt{Cation\_Filter} — Filters by net charge at pH 7. \textit{Inputs}: \texttt{min\_cationicity} (float), 
\texttt{max\_cationicity} (float), \texttt{folder\_path}.

\item \texttt{Hydrophobicity\_Filter} — Filters by GRAVY score, balancing membrane insertion against aqueous solubility. \textit{Inputs}: \texttt{min\_hydrophobicity} (float), \texttt{max\_hydrophobicity} (float), \texttt{folder\_path}.

\item \texttt{Length\_Filter} — Filters by amino acid count. \textit{Inputs}: \texttt{min\_length} (int), \texttt{max\_length} (int), \texttt{folder\_path}.

\item \texttt{Structure\_Filter} — Classifies predicted PDB structures into one of six classes: mixed, $\alpha$-helical, $\beta$-hairpin, extended $\beta$-strand, structured turns, or unstructured. Requires PDB output from \texttt{Predict\_Structure}; does not perform folding itself. The \texttt{structure} input is a semicolon-separated string of class IDs(e.g., \texttt{"2;3"}). \textit{Inputs}: \texttt{folder\_path}, \texttt{structure} (str).

\item \texttt{Damino\_Filter} — Retains sequences containing at least one D-amino acid (lowercase letter in the token stream). Invoked when the user objective specifies non-canonical chemistry. \textit{Inputs}: \texttt{folder\_path}.

\item \texttt{Filter\_Status} — Mandatory bookkeeping tool. Reports the integer count of surviving candidates after each filtering cycle. The Planning Agent uses this output to decide between proceeding, re-filtering, and re-generating.\textit{Inputs}: \texttt{folder\_path}.
\end{itemize}

\paragraph{Verifying Agent.}Cross-references filtered candidates against external databases to assess novelty and biological context. Both BLAST searches return the top hit per candidate within an E-value threshold of 10. While elevated E-value cutoffs are standard for short-peptide BLAST searches, the threshold here is set permissively for demonstration and is exposed as a tunable parameter.

\begin{itemize}
\item \texttt{Verify\_SwissProt} — Local BLAST~\cite{blast} alignment against SwissProt~\cite{Swissprot}, returning top homologs with taxonomic classifications and cross-referenced annotations. \textit{Inputs}: \texttt{folder\_path}.

\item \texttt{Verify\_DBAASP} — BLAST alignment against the DBAASP~\cite{dbaasp} subset used to train AMPGAN v3, returning the closest training-set match together with reported antimicrobial activity. Used to assess training-set proximity and recover known-activity annotations for near-duplicates. \textit{Inputs}: \texttt{folder\_path}.
\end{itemize}

\newpage
\subsection{Representative Trajectory}
\label{app:agentic-trajectory}

We include a representative trajectory from a multi-agent run with objective ``Generate 5  AMPs with D amino acid targeting \textit{E.~coli} with $\alpha$-helical secondary structure.''

\begin{figure*}[ht!]
    \centering
    \includegraphics[width=1.0\textwidth]{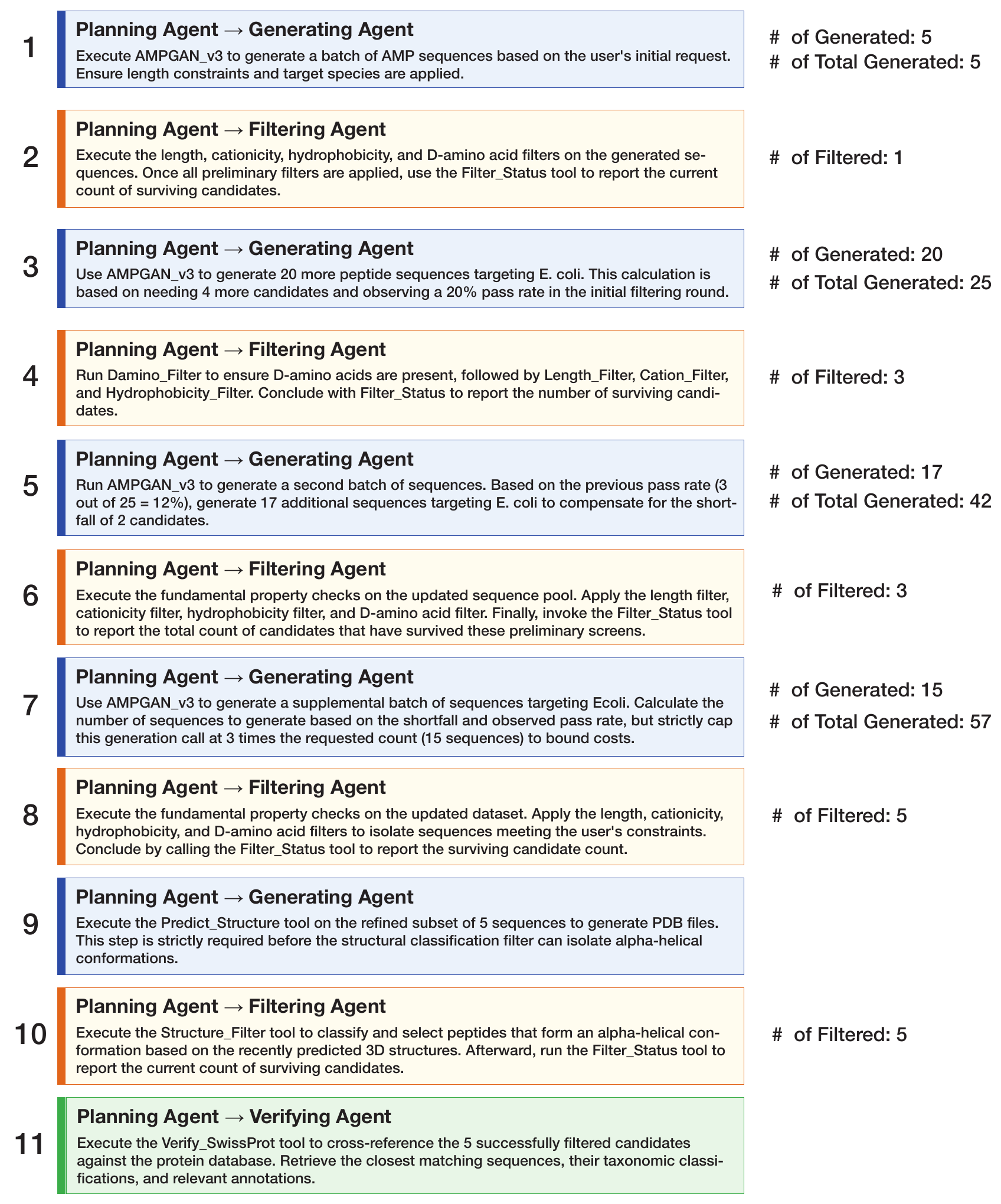}

\end{figure*}




\end{document}